\documentclass[runningheads]{llncs}
\usepackage[T1]{fontenc}
\usepackage{graphicx}
\usepackage{color}
\usepackage{xcolor}
\usepackage{float}
\usepackage{soul}

\begin{document}
\title{Astro, I’m Home! Investigating Factors that Influence the Acceptance of Home Robots Using Supervised Machine Learning}

\author{Katrin Fischer \and
Essence Wilson \and
Steffie Kim \and
Dmitri Williams}
\authorrunning{Fischer et al.}
%
\institute{University of Southern California, Los Angeles, CA 90089, USA \\
\email{katrinfi@usc.edu}} 
%
\maketitle              
\begin{abstract}
The use of social robots in home environments is on the rise. This exploratory study applies regularization techniques (e.g., Lasso and Ridge regression) to investigate variables and identify new models of technology acceptance in the context of social robots. Within the original UTAUT2 framework, performance expectancy, social influence, and hedonic motivation emerged as the strongest and most consistent predictors of intention to use the technology. In addition, usability, trust, and competence were identified as promising variables in a model predicting intention to use.

\end{abstract}

\section{Introduction}

Social robot utilization, receiving broader global recognition with the market size having reached \$5.72 billion \cite{Market}, affords numerous opportunities for humans to obtain quality companionship and emotional support. Indeed, social robots can offer a sense of connection and comfort to users displaced in isolating settings (e.g., elderly patients during the COVID-19 pandemic \cite{jecker2021you}), provide medical support (e.g., assisting aging patients and patients with dementia \cite{chu2017service,gonzalez2021social}), and emotional development \cite{alam2022social,chen2020teaching}. Social robots, too, are advancing computer science scholarship, offering greater affordances for researchers to evaluate user social interactions \cite{deeva2019computational,Wallach2018}. Thus, assessing their potential, including how humans perceive, utilize, and, in turn, accept them, is essential as social robots continue to mobilize in a wide array of sectors to improve human communication strategies.

Traditional acceptance models are commonplace in predicting consumers’ robot utilization and perceptions \cite{salvini2010design}. However, these models fall short in considering robots' embodied forms, as opposed to other digital interfaces (e.g., mobile applications, websites), requiring a reevaluation of additional structures that influence users' adoption behaviors. Thus,  this paper explores additional predictors pertaining to social robot acceptance by leveraging machine learning techniques to predict robot acceptance. This innovative approach holds promise for social robot adoption and offers the possibility of categorizing preferred social and robot attributes that lead to acceptance.

\section{Related Work}

\subsection{Theoretical Framework: Unified Theory of Acceptance and Use of Technology}

Devised by Venkatesh et al. \cite{Venkatesh2012}, the Unified Theory of Acceptance and Use of Technology (UTAUT) is a comprehensive framework to predict technology acceptance across social computer science. Having its roots in eight prior models\footnote{The eight models are Theory of Planned Behavior (TPB; \cite{ajzen1991theory}), Social Cognitive Theory \cite{compeau1999social}, Technology Acceptance Model (TAM; \cite{davis1989perceived}), Motivational Model \cite{davis1992extrinsic}, Theory of Reasoned Action \cite{fishbein1977belief}, Innovation Diffusion Theory \cite{moore1991development}, Combined TAM and TPB \cite{taylor1995understanding}, and Model of PC Utilization (MPCU; \cite{thompson1991personal}).}, UTAUT is widely used in human-computer interaction research \cite{faaeq2013meta,Hajesmaeel-Gohari2022} and has been applied to human-robot interaction contexts \cite{fridin2014acceptance,Rossi2020,Bevilacqua2022,Fischer2025}. Four core factors are (1) Performance expectancy, which can be understood as the extent to which an individual is confident that utilizing the technology will assist them in their professional tasks \cite{davis1992extrinsic,shin2009towards}, (2) effort expectancy, which refers to the degree of ease of using the technology \cite{Venkatesh2003}, (3) facilitating conditions, which are defined as an individual's belief in the presence of supportive organizational infrastructure to eliminate barriers to technology use \cite{thompson1991personal}, and (4) social influence, which reflects an individual's perception of the extent to which their significant peers consider the importance of the technology \cite{diaz2010learning}. 
In addition to these four constructs, Venkatesh et al. \cite{Venkatesh2012} incorporated three additional factors (habit, hedonic motivation, and price value) to extend UTAUT into UTAUT2, which significantly improved the variance explained in behavioral intention and technology use \cite{chang2012utaut}. First, habit is defined as the degree to which individuals engage in actions automatically due to their learned behavior \cite{limayem2007habit}. Next, hedonic motivation refers to the enjoyment experienced when using a technology, which has been demonstrated to be a significant determinant of technology acceptance and use \cite{brown2005model}. Lastly, price value is considered important in this model as the cost structure impacts consumers’ technology use, especially in the consumer use setting instead of the organizational use setting \cite{chang2012utaut}.

\begin{figure}[ht] 
    \centering \includegraphics[width=0.8\textwidth]{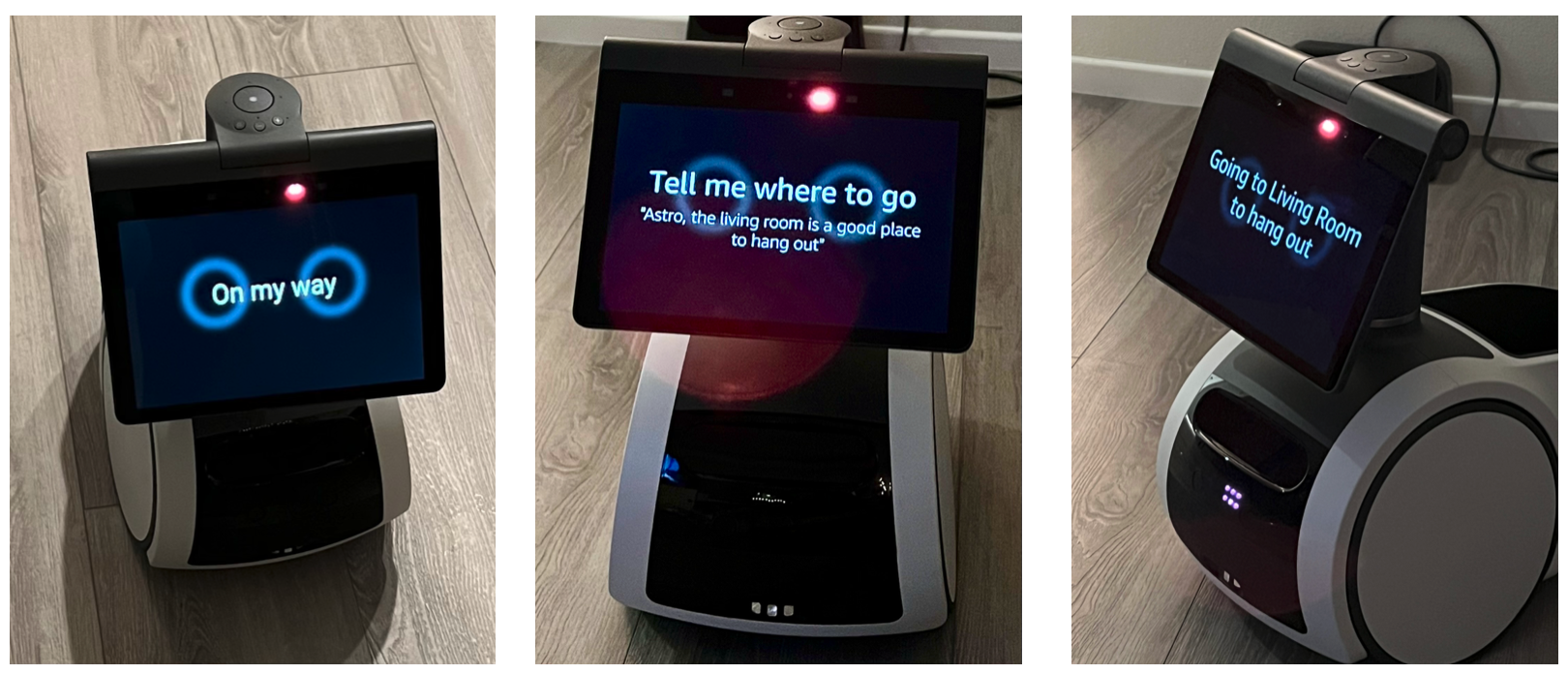} 
    \caption{The Home Robot Astro.}
    \label{fig: Astro Robot}
\end{figure}

\subsection{Potential Predictors of Robot Acceptance}

In addition to established predictors from UTAUT2, this study included six additional variables in the research model: Usability, trust, trustworthiness (ability, benevolence, integrity), warmth, competence, and attitudes toward robots.

First, the relationship between usability and technology acceptance has been well-documented \cite{davis1989perceived,Jung2021}, and the system usability scale \cite{brooke1996sus} is commonly used to assess perceived ease of use in the context of social robots and technology acceptance \cite{Jung2021}. 

Second, the connection between trust and robot acceptance has been well-established \cite{Apraiz2023,Jung2021}. Trust is an attitude \cite{jones1996trust}, and theories informing technology acceptance models are based on the premise that attitudes inform intention, as shown by the theory of reasoned action \cite{fishbein1977belief}, theory of planned behavior \cite{ajzen1991theory}, and TAM \cite{davis1989perceived}. Moreover, research indicates that a robot's trustworthiness characteristics (i.e. ability, benevolence, and integrity), beyond contributing to the development of trust \cite{Kim2020}, directly affect robot acceptance \cite{Fischer2025}. 

Third, people's first impressions are informed by assessments of warmth and competence, which influence the evaluation of upcoming interactions \cite{cuddy2007bias}. Both warmth and competence have been identified as direct antecedents to trustworthiness \cite{Fischer2025} and indirectly affect intention to use social robots \cite{Fischer2023}.

In summary, research has shown that attitudes towards robots are built on the dimensions of utilitarian aspects (ease of use, usefulness), hedonic aspects (emotional attractiveness including appearance, personality, and emotional value of the robot), and trust \cite{Jung2021}, warranting an inclusion of these constructs in considerations of robot acceptance. Based on the literature laid out above, we propose the following research question:

\textbf{RQ:} Which features best predict the intention to use social robots?

\section{Methods}

\subsection{Procedure \& Participants}

\begin{figure}[ht] 
    \centering \includegraphics[width=0.85\textwidth]{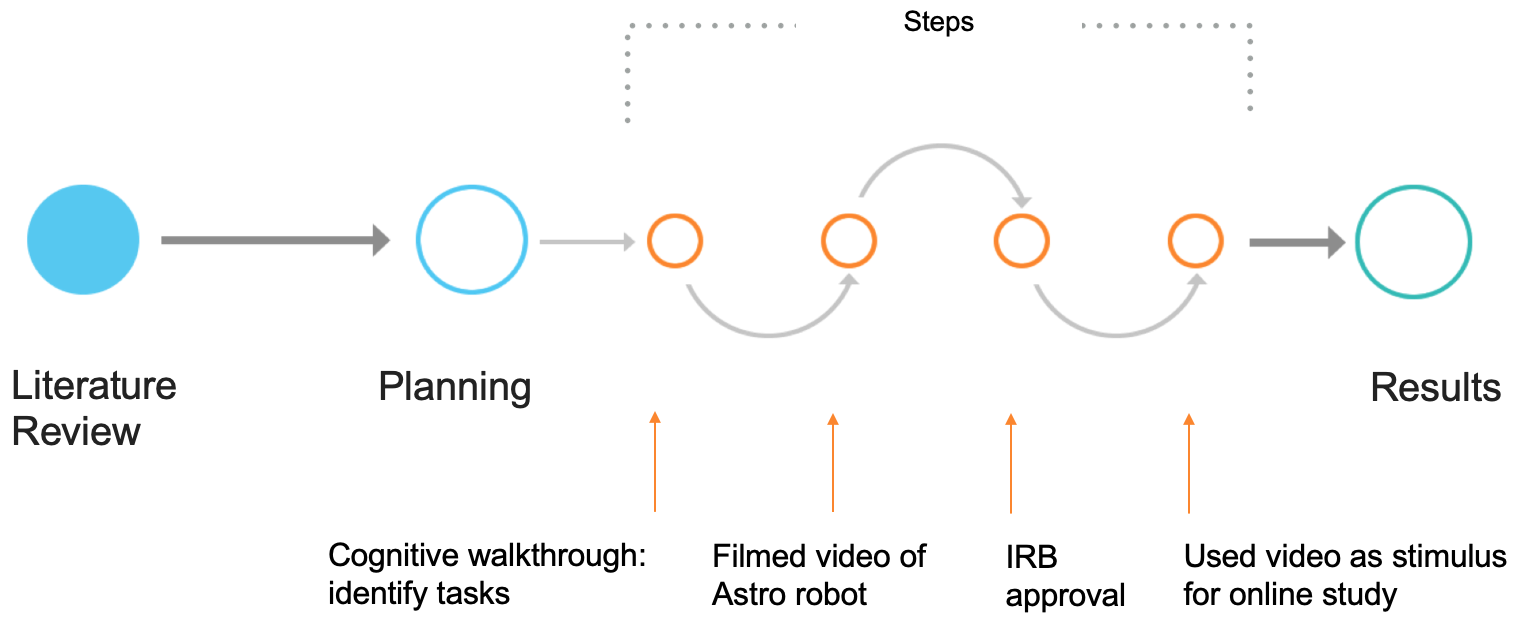} 
    \caption{Study Overview.}
    \label{fig: Study Overview}
\end{figure}

Astro is a social interaction and entertainment service robot \cite{Li2022} that can automatically map, navigate, and monitor homes \cite{Washburn2022}. The Astro robot was deployed to a home environment for two months. A cognitive walkthrough \cite{Lewis1990} was conducted to identify usability barriers pertaining to these tasks. User tasks were identified as navigating the home, interacting with people in the home, and monitoring the home. A video was then filmed depicting Astro completing the tasks and the associated usability challenges and edited to a length of two minutes and 43 seconds. The video stimulus was embedded in a Qualtrics survey and participants were recruited via Amazon Mechanical Turk. The university's IRB approved the study.

\subsection{Measures}

\subsubsection{Demographic Variables.}

Demographic variables included participants' age (measured by birth year), gender (female, male, other), education (in years), race (White/Caucasian, Black African American, Native American, Asian, Native Hawaiian/Pacific Islander, other, prefer not to say), and familiarity with robots (7-point Likert scale).

\subsubsection{UTAUT2 Variables.}

The dependent variable, behavioral intention, and the independent variables of the UTAUT2 model (performance expectancy, effort expectancy, social influence, hedonic motivation, habit, price value) were measured according to Venkatesh et al. \cite{Venkatesh2012}.

\subsubsection{Additional Variables.}

Usability was operationalized via the System Usability Scale (SUS) questionnaire \cite{brooke1996sus} to assess participants’ perceived ease of use of the robot. The latent variables trust and trustworthiness were measured according to \cite{Kim2020}. The first impressions warmth and competence were measured according to the stereotype content model \cite{Fiske2007}.  The NARS scale \cite{Nomura2006} assessed participants’ attitudes towards robots.

\subsection{Data Analysis}

All descriptive and statistical analyses were performed in Python. The general objective of this paper was to identify variables that result in the best prediction for continuous data while avoiding overfitting, and deriving models that are as simple as possible, logical, and well-suited for its purpose. Based on the data available, a supervised model is the best modeling approach, as the outputs are available (labeled). The variables used are the most important factors to successful machine learning projects \cite{domingos2012few}, and this study aims to find those predictors that best explain the outcome (i.e., technology acceptance,  operationalized as an intention to use via the UTAUT model). Removing less relevant features will improve model interpretability and prediction accuracy. 
Thus, the following approaches will be explored: Best subset selection as well as shrinkage or regularization methods (i.e., Lasso and Ridge regression). Subset selection identifies a subset of predictors believed to be best related to the response while shrinkage methods utilize all predictors but use techniques to shrink the estimated coefficients towards zero \cite{james2013introduction}. More specifically, Ridge regression constrains coefficients to make them small but not quite zero (L2 regularization), while Lasso drives coefficients all the way to zero (L1 regularization). The results of these approaches will be compared and provide implications for future technology acceptance models. 

\section{Results}

An online study was run in which 232 participants watched the Astro video and answered questions pertaining to its perceived usability, trust/trustworthiness as well as first impressions and intention to use. After removing 34 participants who failed an attention check or did not complete the full survey, 198 participant responses were available for analysis. Demographic information included participants’ gender (73 F, 125 M), age (\textit{M} = 34.84, \textit{SD} = 9.14), education (\textit{M} = 4.46, \textit{SD} = 3.1), race (89.9\% white/Caucasian), and familiarity with robots (\textit{M} = 5.35, \textit{SD} = 1.36 on a 7-point Likert scale).

\begin{figure}[ht] 
    \centering \includegraphics[width=0.7\textwidth]{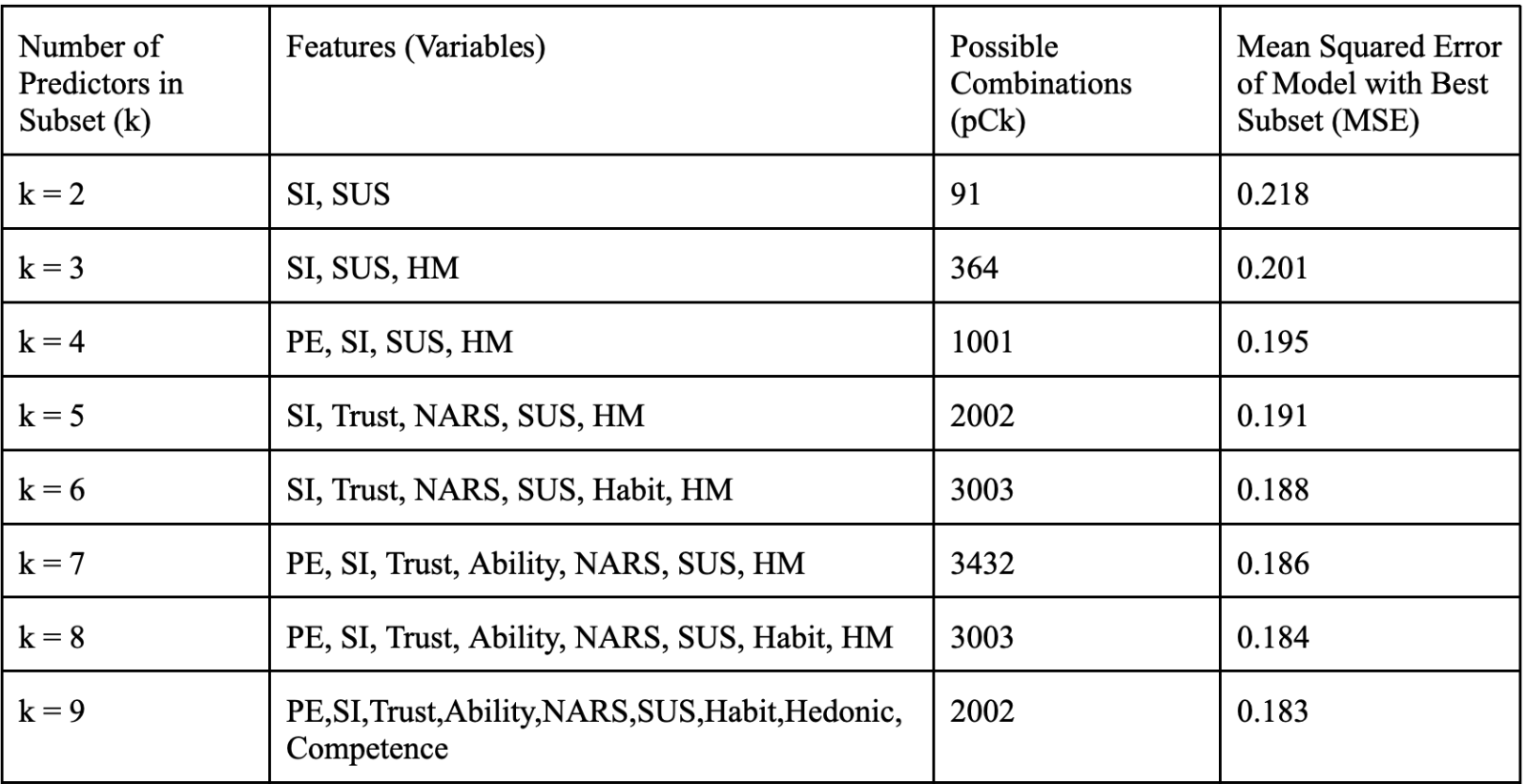} 
    \caption{Best Subset Selection.}
    \label{fig: Best Subset}
\end{figure}

\subsection{Best Subset Selection}

Best subset selection was run and to identify additional variables predicting behavioral intention.
The latent variable behavioral intention ($\alpha$ = 0.79, \textit{M }= 3.88, \textit{SD }= 0.80) was entered as dependent variable. 
Performance expectancy (PE; $\alpha$ =0.69, \textit{M }= 3.91, \textit{SD }= 0.76)  and social influence (SI; $\alpha$ = 0.75, \textit{M }= 3.83, \textit{SD }= 0.76) are the strongest predictors from the original UTAUT model. Participants’ perceptions of the  usability  of the robot (SUS; $\alpha$ = 0.74, \textit{M }= 3.98, \textit{SD }= 0.63) and  trust ($\alpha$ = 0.91, \textit{M }= 5.32, \textit{SD }= 1.06) as well as the trustworthiness sub dimension ability ($\alpha$ = 0.89, \textit{M }= 5.51, \textit{SD }= 1.03) and negative attitudes (NARS; $\alpha$ = 0.78, \textit{M }= 3.62, \textit{SD }= 0.59) emerge as important predictors (see Figure \ref{fig: Best Subset}).

\subsection{Lasso Regression}
\begin{figure}[ht] 
    \centering \includegraphics[width = 1\textwidth]{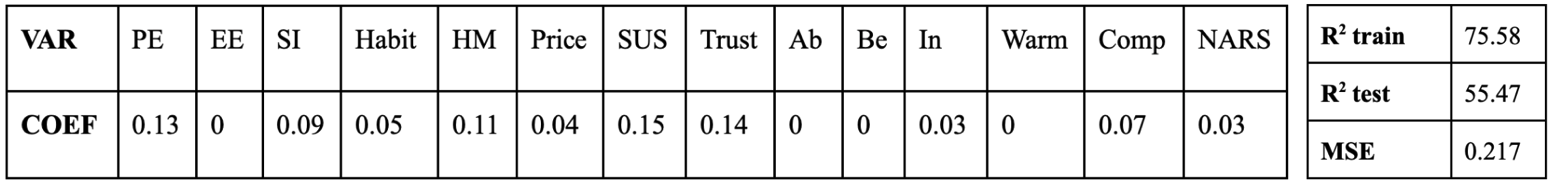} 
    \caption{Lasso Regression Results.}
    \label{fig: Lasso}
\end{figure}

In preparation for Lasso regression, the dataset was split into a train (70\%) and test (30\%) set. All features were preprocessed to center around zero. An array of possible lambdas (alphas) was generated to fit 100 models. To find its optimal value, we used sklearn’s Lasso linear model with iterative fitting along a regularization path. The best model, with the best value of penalization (lambda $\lambda$ = 0.03), was selected by 5-fold cross-validation. The Lasso model including the best lambda was run and drove several coefficients to zero (see Figure \ref{fig: Lasso}).

In the best model selected by Lasso regression, performance expectancy (PE), social influence (SI), and hedonic motivation (HM; $\alpha$ = 0.70, \textit{M }= 4.03, \textit{SD }= 0.69) are the strongest UTAUT variables. Usability and trust are the strongest predictors among the added variables of interest. The coefficients for effort expectancy (EE; $\alpha$ = 0.68, \textit{M }= 4.03, \textit{SD }= 0.60), ability (Ab), benevolence (Be; $\alpha$ = 0.86, \textit{M }= 5.40, \textit{SD }= 1.04) and the first impression of warmth ($\alpha$ = 0.76, \textit{M }= 5.54, \textit{SD }= 0.97) have been driven to zero.

\subsection{Ridge Regression}

\begin{figure}[ht] 
    \centering \includegraphics[width = 1\textwidth]{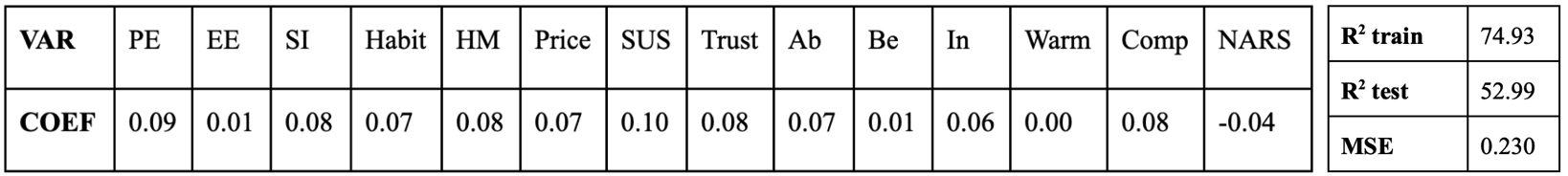} 
    \caption{Ridge Regression Results.}
    \label{fig: Ridge}
\end{figure}

Performance expectancy (PE), social influence (SI) and hedonic motivation (HM) are three strongest UTAUT predictors (closely followed by habit and price). Usability, trust and  competence (Comp; $\alpha$ = 0.83, \textit{M }= 5.52, \textit{SD }= 1.02) are the strongest additional variables. The coefficients for benevolence and warmth have shrunk the most.

Thus, all models retained performance expectancy, social influence and hedonic motivation from the original UTAUT2 model. Usability and trust were the strongest additional predictors, while some approaches also highlight robot competence and ability.

\section{Discussion}

Social robots continue to receive global recognition in market spaces \cite{Market} and across numerous social settings (e.g., in-home services, clinical and outpatient support, youth socioemotional learning \cite{chu2017service,gonzalez2021social,alam2022social,chen2020teaching}). 
However, traditional acceptance models (e.g., UTAUT) have historically failed to account for the embodied features of social robots, having primarily been validated with their digital interfaces. As such, more research is required to match the speed of its growth.

This exploratory study used supervised machine learning, including regularization methods (e.g., Lasso and Ridge), to investigate variable selection integrity from UTAUT and discover new models of technology acceptance in the context of social robots. Of the original model, the variables performance expectancy, social influence, and hedonic motivation emerged as the strongest and consistent predictors of intention to use robots, in line with research investigating technology use in other contexts \cite{Venkatesh2012,chang2012utaut}. Performance expectancy specifically has been identified as the most significant contributor in prior research \cite{chang2012utaut}. Usability, trust, and competence or ability also emerged as promising variables in a model predicting the outcome of interest. These results contribute to the growing body of literature recognizing the important role of trust when it comes to human-robot interactions \cite{Nam2020,hancock2011meta,Naneva2020,salem2015evaluating,Salem2015} as well as the robot's agentic capabilities \cite{glawe2025human,heerink2006influence}. Moreover, usability, or ease of use, has increasingly influenced technology use as indicated by the growth of user experience assessments for human-computer interactions \cite{Hassenzahl2006} as well as calls for its integrations when it comes to social robots \cite{Shourmasti2021,alenljung2015user}.

Regularization, a widely used and accepted supervised machine learning technique for variable selection \cite{james2013introduction}, was selected as the core method to shape acceptance models specific for social robots. Lasso identified the most important features by shrinking poor fitting variables to 0. Meanwhile, Ridge handles violations of multicollinearity by shrinking and thus stabilizing each variables' coefficients without necessarily eliminating features. Both regularization methods were tested and triangulated to confirm robust patterns on our stand-out variables while controlling for statistical assumptions.

These results give call for future models to incorporate these top performing variables to assess robot acceptance including motion to develop more scales. Given the growing ecosystem of embodied artificial intelligence systems, existing measurement instruments may have to be adjusted for social robot contexts. For instance, usability was measured by use of the system usability scale \cite{brooke1996sus}, an instrument developed in the context of computers in the workplace, and has not been re-assessed for social robots. Future research is currently underway to identify additional dimensions of robot usability, whose assessment contributes to the explainability of ease of use in this context. In follow-up to this study, future research will aim to collect more data to build and test one comprehensive model, validated by structural equation modeling methods and robust evaluation metrics. 

This research has a bearing on our understanding of the factors involved in people’s acceptance of social robots. 
Predicting consumers’ intention to use, this study employs validated, theory-driven measures to examine how relevant variables emerge in the context of social robots and provides leeway to further develop acceptance models under integration of insights into new measurement instruments.

%
%
\bibliographystyle{splncs04}
\bibliography{References}

@article{Wallach2018,
   author = {Hanna Wallach},
   doi = {10.1145/3132698},
   issn = {15577317},
   issue = {3},
   journal = {Communications of the ACM},
   pages = {42-44},
   title = {Viewpoint: computational social science $\neq$ computer science + social data},
   volume = {61},
   year = {2018}
}

@article{salvini2010design,
  title={Design for acceptability: improving robots’ coexistence in human society},
  author={Salvini, Pericle and Laschi, Cecilia and Dario, Paolo},
  journal={International journal of social robotics},
  volume={2},
  number={4},
  pages={451--460},
  year={2010},
  publisher={Springer}
}

@article{Venkatesh2012,
   author = {Viswanath Venkatesh and James Y.L. Thong and Xin Xu},
   doi = {10.2307/41410412},
   issue = {1},
   journal = {MIS Quarterly: Management Information Systems},
   pages = {157-178},
   title = {Consumer acceptance and use of information technology: Extending the unified theory of acceptance and use of technology},
   volume = {36},
   year = {2012}
}

@article{Hajesmaeel-Gohari2022,
    author = {Sadrieh Hajesmaeel-Gohari and Firoozeh Khordastan and Farhad Fatehi and Hamidreza Samzadeh and Kambiz Bahaadinbeigy},
   doi = {10.1186/s12911-022-01764-2},
   issue = {1},
   journal = {BMC Medical Informatics and Decision Making},
   pages = {1-9},
   
   publisher = {BioMed Central},
   title = {The most used questionnaires for evaluating satisfaction, usability, acceptance, and quality outcomes of mobile health},
   volume = {22},
   
   year = {2022}
}

@article{fridin2014acceptance,
   author = {Marina Fridin and Mark Belokopytov},
   journal = {Computers in Human Behavior},
   pages = {23-31},
   publisher = {Elsevier},
   title = {Acceptance of socially assistive humanoid robot by preschool and elementary school teachers},
   volume = {33},
   year = {2014}
}

@article{Rossi2020,
   author = {S Rossi and D Conti and F Garramone and G Santangelo and M Staffa and S Varrasi and A Di Nuovo},
   doi = {10.3390/robotics9020039},
   issue = {2},
   journal = {Robotics},
   pages = {1-19},
   title = {The role of personality factors and empathy in the acceptance and performance of a social robot for psychometric evaluations},
   volume = {9},
   year = {2020}
}

@article{Bevilacqua2022,
   author = {Roberta Bevilacqua and Mirko Di Rosa and Giovanni Renato Riccardi and Giuseppe Pelliccioni and Fabrizia Lattanzio and Elisa Felici and Arianna Margaritini and Giulio Amabili and Elvira Maranesi},
   doi = {10.3389/fnbot.2022.883106},
   issue = {July},
   journal = {Frontiers in Neurorobotics},
   pages = {1-11},
   title = {Design and development of a scale for valuating the acceptance of social robotics for older people: The robot era inventory},
   volume = {16},
   year = {2022}
}

@article{Venkatesh2003,
   author = {Viswanath Venkatesh and Michael G. Morris and Gordon B. Davis and Fred D. Davis},
   doi = {10.2307/30036540},
   issue = {3},
   journal = {MIS Quarterly: Management Information Systems},
   pages = {425-478},
   publisher = {Management Information Systems Research Center},
   title = {User acceptance of information technology: Toward a unified view},
   volume = {27},
   year = {2003}
}

@article{jecker2021you,
  title={You've got a friend in me: sociable robots for older adults in an age of global pandemics},
  author={Jecker, Nancy S},
  journal={Ethics and Information Technology},
  volume={23},
  number={Suppl 1},
  pages={35--43},
  year={2021},
  publisher={Springer}
}

@article{taylor1995understanding,
  title={Understanding information technology usage: A test of competing models},
  author={Taylor, Shirley and Todd, Peter A},
  journal={Information systems research},
  volume={6},
  number={2},
  pages={144--176},
  year={1995},
  publisher={INFORMS}
}

@article{chu2017service,
  title={Service innovation through social robot engagement to improve dementia care quality},
  author={Chu, Mei-Tai and Khosla, Rajiv and Khaksar, Seyed Mohammad Sadegh and Nguyen, Khanh},
  journal={Assistive Technology},
  volume={29},
  number={1},
  pages={8--18},
  year={2017},
  publisher={Taylor \& Francis}
}

@inproceedings{Washburn2022,
   author = {Andrew Washburn and Kripash Shrestha and Habib Ahmed and Dave Feil-Seifer and Hung Manh La},
   doi = {10.1109/ICHMS56717.2022.9980806},
   booktitle = {Proceedings of the 2022 IEEE International Conference on Human-Machine Systems, ICHMS 2022},
   pages = {1-6},
   publisher = {IEEE},
   title = {Exploring human compliance toward a package delivery robot},
   year = {2022}
}

@article{Nomura2006,
   author = {Tatsuya Nomura and Takayuki Kanda and Tomohiro Suzuki},
   doi = {10.1007/s00146-005-0012-7},
   issue = {2},
   journal = {AI and Society},
   month = {3},
   pages = {138-150},
   publisher = {Springer},
   title = {Experimental investigation into influence of negative attitudes toward robots on human-robot interaction},
   volume = {20},
   year = {2006}
}

@article{moore1991development,
  title={Development of an instrument to measure the perceptions of adopting an information technology innovation},
  author={Moore, Gary C and Benbasat, Izak},
  journal={Information systems research},
  volume={2},
  number={3},
  pages={192--222},
  year={1991},
  publisher={Informs}
}

@article{shin2009towards,
  title={Towards an understanding of the consumer acceptance of mobile wallet},
  author={Shin, Dong-Hee},
  journal={Computers in human behavior},
  volume={25},
  number={6},
  pages={1343--1354},
  year={2009},
  publisher={Elsevier},
  doi = {10.1016/j.chb.2009.06.001}
}

@article{thompson1991personal,
  title={Personal computing: Toward a conceptual model of utilization},
  author={Thompson, Ronald L and Higgins, Christopher A and Howell, Jane M},
  journal={MIS quarterly},
  pages={125--143},
  year={1991},
  publisher={JSTOR}
}

@article{gonzalez2021social,
  title={Social robots in hospitals: a systematic review},
  author={Gonz{\'a}lez-Gonz{\'a}lez, Carina Soledad and Violant-Holz, Ver{\'o}nica and Gil-Iranzo, Rosa Maria},
  journal={Applied Sciences},
  volume={11},
  number={13},
  pages={5976},
  year={2021},
  publisher={MDPI}
}

@article{alam2022social,
  title={Social robots in education for long-term human-robot interaction: socially supportive behaviour of robotic tutor for creating robo-tangible learning environment in a guided discovery learning interaction},
  author={Alam, Ashraf},
  journal={ECS Transactions},
  volume={107},
  number={1},
  pages={12389},
  year={2022},
  publisher={IOP Publishing}
}

@book{Nam2020,
editor = {Nam, CS and Lyons, JB},
publisher = {Elsevier},
title = {{Trust in human-robot interaction: Research and applications}},
year = {2020}
}

@article{Naneva2020,
author = {Naneva, Stanislava and {Sarda Gou}, Marina and Webb, Thomas L. and Prescott, Tony J.},
doi = {10.1007/s12369-020-00659-4},
journal = {International Journal of Social Robotics},
month = {dec},
pages = {1179--1201},
publisher = {Springer Science and Business Media B.V.},
title = {{A systematic review of attitudes, anxiety, acceptance, and trust towards social robots}},
volume = {12},
year = {2020}
}

@inproceedings{heerink2006influence,
   author = {Marcel Heerink and Ben Krose and Vanessa Evers and Bob Wielinga},
   institution = {IEEE},
   booktitle = {ROMAN 2006-The 15th IEEE International Symposium on Robot and Human Interactive Communication},
   pages = {521-526},
   title = {The influence of a robot's social abilities on acceptance by elderly users},
   year = {2006}
}

@article{Shourmasti2021,
   author = {Elaheh Shahmir Shourmasti and Ricardo Colomo-Palacios and Harald Holone and Selina Demi},
   doi = {10.3390/s21155052},
   issn = {14248220},
   issue = {15},
   journal = {Sensors},
   pages = {1-19},
   pmid = {34372289},
   title = {User experience in social robots},
   volume = {21},
   year = {2021}
}

@article{glawe2025human,
  title={Human Autonomy and Sense of Agency in Human-Robot Interaction: A Systematic Literature Review},
  author={Glawe, Felix and Schmeckel, Tim and Brauner, Philipp and Ziefle, Martina},
  journal={arXiv preprint arXiv:2509.22271},
  year={2025}
}

@article{Hassenzahl2006,
   
   author = {Marc Hassenzahl and Noam Tractinsky},
   doi = {10.1080/01449290500330331},
   isbn = {0144929050033},
   issn = {0144929X},
   issue = {2},
   journal = {Behaviour and Information Technology},
   pages = {91-97},
   title = {User experience - A research agenda},
   volume = {25},
   year = {2006}
}

@inbook{alenljung2015user,
   author = {Beatrice Alenljung and Jessica Lindblom},
   booktitle = {Handbook of Research on Synthesizing Human Emotion in Intelligent Systems and Robotics},
   pages = {352-364},
   publisher = {IGI Global},
   title = {User experience of socially interactive robots: its role and relevance},
   year = {2015}
}

@article{salem2015evaluating,
author = {Salem, Maha and Dautenhahn, Kerstin},
journal = {Emerging Policy and Ethics of Human-Robot Interaction},
publisher = {ACM Press},
title = {{Evaluating trust and safety in HRI: Practical issues and ethical challenges}},
year = {2015}
}

@inproceedings{Salem2015,
author = {Salem, Maha and Lakatos, Gabriella and Amirabdollahian, Farshid and Dautenhahn, Kerstin},
booktitle = {Proceedings of the Tenth Annual ACM/IEEE International Conference on Human-Robot Interaction},
doi = {10.1145/2696454.2696497},
pages = {141--148},
publisher = {ACM},
title = {{Would you trust a (faulty) robot?: Effects of error, task type and personality on human-robot cooperation and trust}},
year = {2015}
}

@article{hancock2011meta,
author = {Hancock, Peter A and Billings, Deborah R and Schaefer, Kristin E and Chen, Jessie Y C and {De Visser}, Ewart J and Parasuraman, Raja},
journal = {Human factors},
number = {5},
pages = {517--527},
publisher = {Sage Publications Sage CA: Los Angeles, CA},
title = {{A meta-analysis of factors affecting trust in human-robot interaction}},
volume = {53},
year = {2011}
}

@article{ajzen1991theory,
  title={The theory of planned behavior},
  author={Ajzen, Icek},
  journal={Organizational behavior and human decision processes},
  volume={50},
  number={2},
  pages={179--211},
  year={1991},
  publisher={Elsevier}
}

@article{Apraiz2023,
   author = {Ainhoa Apraiz and Ganix Lasa and Maitane Mazmela},
   doi = {10.1007/s12369-022-00957-z},
   issue = {2},
   journal = {International Journal of Social Robotics},
   pages = {187-210},
   publisher = {Springer Netherlands},
   title = {Evaluation of user experience in human–robot interaction: A systematic literature review},
   volume = {15},
   year = {2023}
}

@article{brooke1996sus,
  title={SUS-A quick and dirty usability scale},
  author={Brooke, John and others},
  journal={Usability evaluation in industry},
  volume={189},
  number={194},
  pages={4--7},
  year={1996},
  publisher={London, England}
}

@article{brown2005model,
  title={Model of adoption of technology in households: A baseline model test and extension incorporating household life cycle},
  author={Brown, Susan A and Venkatesh, Viswanath},
  journal={MIS quarterly},
  pages={399--426},
  year={2005},
  publisher={JSTOR}
}

@article{chang2012utaut,
  title={UTAUT and UTAUT 2: A review and agenda for future research},
  author={Chang, Andreas},
  journal={Journal the WINNERS},
  volume={13},
  number={2},
  pages={10--114},
  year={2012}
}

@inproceedings{Fischer2025,
   author = {Katrin Fischer and Donggyu Kim and Joo-Wha Hong},
   doi = {10.1007/978-3-031-93412-4_4},
   editor = {H. Degen and S. Ntoa},
   booktitle = {Artificial Intelligence in HCI. HCII 2025. Lecture Notes in Computer Science},
   pages = {67-85},
   publisher = {Springer, Cham.},
   title = {What makes people use social robots? {I}ntegrating trustworthiness into the {UTAUT} model},
   volume = {15819},
   year = {2025}
}

@inproceedings{Fischer2023,
   author = {Katrin Fischer and Donggyu Kim and Joo-Wha Hong},
   doi = {10.48550/arXiv.2311.06688},
   booktitle = {32nd IEEE International Conference on Robot \& Human Interactive Communication (RO-MAN), SCRITA Workshop on Trust, Acceptance and Social Cues in Human-Robot Interaction},
   title = {The effect of trust and its antecedents on robot acceptance},
   year = {2023}
}

@article{chen2020teaching,
  title={Teaching and learning with children: Impact of reciprocal peer learning with a social robot on children’s learning and emotive engagement},
  author={Chen, Huili and Park, Hae Won and Breazeal, Cynthia},
  journal={Computers \& Education},
  volume={150},
  pages={103836},
  year={2020},
  publisher={Elsevier}
}

@article{compeau1999social,
  title={Social cognitive theory and individual reactions to computing technology: A longitudinal study},
  author={Compeau, Deborah and Higgins, Christopher A and Huff, Sid},
  journal={MIS quarterly},
  pages={145--158},
  year={1999},
  publisher={JSTOR}
}

@article{cuddy2007bias,
  title={The BIAS map: behaviors from intergroup affect and stereotypes.},
  author={Cuddy, Amy JC and Fiske, Susan T and Glick, Peter},
  journal={Journal of personality and social psychology},
  volume={92},
  number={4},
  pages={631},
  year={2007},
  publisher={American Psychological Association},
  doi = {10.1037/0022-3514.92.4.631}
}

@article{davis1989perceived,
  title={Perceived usefulness, perceived ease of use, and user acceptance of information technology},
  author={Davis, Fred D},
  journal={MIS quarterly},
  pages={319--340},
  year={1989},
  publisher={JSTOR},
  doi = {10.2307/249008}
}

@article{davis1992extrinsic,
  title={Extrinsic and intrinsic motivation to use computers in the workplace 1},
  author={Davis, Fred D and Bagozzi, Richard P and Warshaw, Paul R},
  journal={Journal of applied social psychology},
  volume={22},
  number={14},
  pages={1111--1132},
  year={1992},
  publisher={Wiley Online Library}
}

@article{deeva2019computational,
  title={Computational personality prediction based on digital footprint of a social media user},
  author={Deeva, Irina},
  journal={Procedia computer science},
  volume={156},
  pages={185--193},
  year={2019},
  publisher={Elsevier}
}

@article{diaz2010learning,
  title={Learning new uses of technology while on an audit engagement: Contextualizing general models to advance pragmatic understanding},
  author={Diaz, Michelle Chandler and Loraas, Tina},
  journal={International Journal of Accounting Information Systems},
  volume={11},
  number={1},
  pages={61--77},
  year={2010},
  publisher={Elsevier},
  doi = {10.1016/j.accinf.2009.05.001}
}

@article{domingos2012few,
  title={A few useful things to know about machine learning},
  author={Domingos, Pedro},
  journal={Communications of the ACM},
  volume={55},
  number={10},
  pages={78--87},
  year={2012},
  publisher={ACM New York, NY, USA}
}

@article{faaeq2013meta,
  title={A meta--analysis of the unified theory of acceptance and use of technology studies among several countries},
  author={Faaeq, Munadil K and Ismail, Noor Azizi and Osman, Wan Rozaini Sheik and Al--Swidi, Abdullah Kaid and Faieq, Alaa K},
  journal={Electronic Government, an International Journal},
  volume={10},
  number={3-4},
  pages={343--360},
  year={2013},
  publisher={Inderscience Publishers}
}

@article{fishbein1977belief,
  title={Belief, attitude, intention, and behavior: An introduction to theory and research},
  author={Fishbein, Martin and Ajzen, Icek},
  journal={Philosophy and Rhetoric},
  year={1977}
}

@article{Fiske2007,
   author = {Susan T. Fiske and Amy J.C. Cuddy and Peter Glick},
   doi = {10.1016/J.TICS.2006.11.005},
   issue = {2},
   journal = {Trends in Cognitive Sciences},
   month = {2},
   pages = {77-83},
   publisher = {Elsevier Current Trends},
   title = {Universal dimensions of social cognition: Warmth and competence},
   volume = {11},
   year = {2007}
}

@book{james2013introduction,
  title={An introduction to statistical learning: with applications in R},
  author={James, Gareth and Witten, Daniela and Hastie, Trevor and Tibshirani, Robert},
  year={2013},
  publisher={Springer}
}

@article{jones1996trust,
  title={Trust as an affective attitude},
  author={Jones, Karen},
  journal={Ethics},
  volume={107},
  number={1},
  pages={4--25},
  year={1996},
  publisher={University of Chicago Press}
}

@article{Jung2021,
   author = {M Jung and M J S Lazaro and M H Yun},
   doi = {10.3390/app11041388},
   issue = {4},
   journal = {Applied Sciences},
   pages = {1-18},
   title = {Evaluation of methodologies and measures on the usability of social robots: A systematic review},
   volume = {11},
   year = {2021}
}

@article{Kim2020,
   author = {Wonjoon Kim and Nayoung Kim and Joseph B. Lyons and Chang S. Nam},
   doi = {10.1016/j.apergo.2020.103056},
   journal = {Applied Ergonomics},
   month = {5},
   publisher = {Elsevier Ltd},
   title = {Factors affecting trust in high-vulnerability human-robot interaction contexts: A structural equation modelling approach},
   volume = {85},
   year = {2020}
}

@article{Lewis1990,
    author = {Clayton Lewis and Peter Poison and Cathleen Wharton and John Rieman},
   doi = {10.1145/97243.97279},
   journal = {Conference on Human Factors in Computing Systems - Proceedings},
   pages = {235-242},
   title = {Testing a Walkthrough Methodology for Theory-Based Design of Walk-Up-and-Use Interfaces},
   year = {1990}
}

@inproceedings{Li2022,
   author = {Shuhai Li and Yuqi Liu},
   doi = {10.1109/IoTaIS56727.2022.9976005},
   booktitle = {Proceedings of the 2022 IEEE International Conference on Internet of Things and Intelligence Systems (IoTaIS)},
   pages = {395-401},
   publisher = {IEEE},
   title = {How can smart service robot help the elderly aging in place: Application, prospect and preference},
   year = {2022}
}

@article{limayem2007habit,
  title={How habit limits the predictive power of intention: The case of information systems continuance},
  author={Limayem, Moez and Hirt, Sabine Gabriele and Cheung, Christy MK},
  journal={MIS quarterly},
  pages={705--737},
  year={2007},
  publisher={JSTOR}
}

@misc{Market,
    author = {{The Business Research Company}},
   title = {Social Robots Global Market Report},
   url = {https://www.thebusinessresearchcompany.com/report/social-robots-global-market-report},
   year = {2025}
}

\end{document}